\documentclass[11pt,a4paper]{article}
\pdfoutput=1
\usepackage{jcappub}

\usepackage{verbatim}
\usepackage{amsmath, multirow, amssymb, wasysym, gensymb}
\usepackage{dcolumn}   
\usepackage{bm}        
\usepackage{amssymb}   
\usepackage{amsmath}
\renewcommand {\t}{\tilde{\theta}}
\usepackage{graphicx}
\usepackage{amsfonts}
\usepackage{graphicx}
\usepackage{graphics}
\usepackage[ngerman, english]{babel}
\usepackage{float}
\usepackage{youngtab}
\usepackage{subfigure}
\usepackage[utf8]{inputenc}
 \usepackage{mathrsfs}

\begin{document}
\title{Flavor ratios of extragalactic neutrinos and neutrino shortcuts in extra dimensions}
\author[a]{Elke Aeikens,}
\author[a]{Heinrich P\"as,}
\author[b]{Sandip Pakvasa,}
\author[a]{Philipp Sicking}
\affiliation[a]{Fakult\"at f\"ur Physik,
Technische Universit\"at Dortmund, 44221 Dortmund,
Germany}
\affiliation[b]{
Department of Physics \& Astronomy, University of Hawaii, 
Honolulu, HI 96822, USA}

\emailAdd{elke.aeikens@tu-dortmund.de}
\emailAdd{heinrich.paes@tu-dortmund.de}
\emailAdd{pakvasa@phys.hawaii.edu}
\emailAdd{philipp.sicking@tu-dortmund.de}
\abstract{
The recent measurement of high energy extragalactic neutrinos by the IceCube Collaboration 
has opened a new window to
probe non-standard neutrino properties. Among other effects, sterile neutrino altered dispersion relations (ADRs) due to shortcuts in an extra
dimension can significantly affect astrophysical flavor ratios. We discuss two limiting cases of this effect, first active-sterile neutrino
oscillations with a constant  ADR potential and second an MSW-like resonant conversion arising from geodesics oscillating
around the brane in an asymmetrically warped extra dimension. We demonstrate that the second case has the potential to suppress
significantly the flux of specific flavors such as $\nu_\mu$ or $\nu_\tau$ at high energies.}


\arxivnumber{1410.0408}

\maketitle
\flushbottom
\section{Introduction}
Recently the IceCube Collaboration has reported the detection of 37 neutrino events in the energy range of 30 TeV up to 2 PeV.
Atmospheric neutrinos as the source of this signal are ruled out at more than 5.7 sigma \cite{Aartsen:2014gkd} 
and it has been argued that these neutrinos are extragalactic in origin
\cite{Aartsen:2013jdh}.
Consequently this result allows for the first time to study the flavor composition of  high energy extra-galactic neutrinos and thus opens up
a window onto neutrino properties exhibiting themselves at extreme energies and propagation distances. 
An apparent deficit of muon track events in the IceCube data has been discussed as a possible 
indication for new physics beyond the Standard Model 
\cite{Pakvasa:2012db,Dorame:2013lka,Mena:2014sja}, albeit this anomaly is so far not statistically significant
\cite{Chen:2013dza}.
In spite of the seeming deficit of muons, several flavor analyses of IceCube events find that the data are consistent with the expected canonical mix of 1:1:1 \cite{Anchordoqui:2013dnh,Palomares-Ruiz:2015mka,Palladino:2015zua,Aartsen:2015ivb}.
On the other hand a recent spectral analysis \cite{Palomares-Ruiz:2015mka} points towards an energy dependent flavor composition
with muon and tau neutrino depletion at high energies while the analysis \cite{Aartsen:2015ivb} prefers a best fit flavor ratio of 0:0.2:0.8, both of 
which are very hard to understand without invoking new physics. 
However, more statistics is needed to determine the flavor ratio of extra-galactic neutrinos and a deviation from 1:1:1 to probe any signatures 
of such new physics.
\section{ADR due to sterile neutrino shortcuts}
In this paper we discuss
how scenarios in which the simple dispersion relation $E^2 = p^2 + m^2$ is altered due to sterile neutrino shortcuts in extra dimensions 
\cite{Pas:2005rb}
can affect these flavor ratios. For example, sterile neutrinos propagating in  
spacetimes with an asymmetrically warped extra dimension $u$
\cite{Chung:1999xg,Chung:1999zs,Csaki:2000dm}
\begin{eqnarray}
         ds^2 = dt^2 - \sum_{i=1}^{3} \eta^{2}(u) \left(dx^i\right)^2 - du^2 \label{eq_metric}
     \end{eqnarray}
 propagate on geodesics oscillating around the brane \cite{Hollenberg:2009ws}
and experience a shorter propagation time \cite{Pas:2005rb} which can be
 parametrized by a shortcut parameter defined as the normalized 
difference of propagation times on the brane and in the bulk, 
$\epsilon \equiv \delta t/t$. 
While being speculative, such scenarios where sterile neutrinos acquire 
altered dispersion relations are motivated by several anomalies in
neutrino oscillations experiments and cosmology, for an overview see 
\cite{Kopp:2013vaa}.

As a consequence, the effective neutrino masses and mixing are
altered in a way similar to what happens when neutrinos propagate inside matter
\cite{Pas:2005rb}, so that the effective two-flavor oscillation amplitude becomes
\begin{eqnarray}
\sin^2 2\tilde{\theta}  = 
\frac{\sin^2 2\theta}{\sin^2 2\theta +\cos^2 2\theta 
	\left[1-\left(\frac{E}{E_{\rm Res}}\right)^2\right]^2}\,, \label{eq_sinq}
\end{eqnarray}
and the Hamiltonian is altered by the additional term
\begin{eqnarray}
\delta \tilde{H} = \frac{\delta m^2}{2E} \sqrt{\sin^2 2 \theta + \cos^2 2 \theta \left(1-\left(\frac{E}{E_{\rm Res}}\right)^2\right)},
\end{eqnarray}
where  $E_{\rm Res}$ denotes the resonance energy 
\begin{eqnarray}
E_{\rm Res} = \sqrt{\frac{\delta m^2 \cos 2\theta}{2\,\epsilon}}\,.
\end{eqnarray}

Note that for energies much smaller than the resonance energy $E_{\rm Res}$, 
the change in the dispersion relation due to the extra-dimensional shortcut
decouples and the scenario discussed resembles the standard scenario with 
three active and one sterile neutrino.

Further details of such models have been  worked out in 
\cite{Pas:2006si,Dent:2007rk,Hollenberg:2009bq,Marfatia:2011bw} and
a similar effect due to the interaction with the matter potential of intergalactic dark matter has been discussed in  \cite{Miranda:2013wla}.
\section{Constant ADR potential}
Astrophysical neutrinos are typically assumed to originate from a pion source  
where the dominant number of neutrinos is produced in the decays of charged pions and kaons resulting
from high energy proton-proton and few proton-photon collisions \cite{Beacom:2003nh,Lai:2009ke}, 
$p+p\rightarrow\pi^\pm\rightarrow\mu^\pm
+\overset{(-)}{\nu_\mu}\rightarrow e^\pm+\overset{(-)}{\nu_e}+\nu_\mu+\bar{\nu}_\mu$. 
In this case the initial flavor ratio at the source is 
\begin{equation}
(\Phi_e^0:\Phi_\mu^0:\Phi_\tau^0)=(1:2:0)\label{3ad},
\end{equation}
where $\Phi_\alpha$ denotes the integrated flux of both neutrinos and anti-neutrinos.
To estimate the flavor ratios after propagation from the source to the detector we follow \cite{Pakvasa:2007dc}.

As neutrino oscillations decohere over the large propagation distances the flavor conversion 
is properly described by the mean of the oscillation probability so that
the oscillatory term averages out to $\sin^2(x/L_{osz})\sim 1/2$ 
and the conversion probability $P_{\alpha\beta}$ and consequently the normalized neutrino fluxes
depend only on the mixing matrix elements,
\begin{eqnarray}
 \Phi_\alpha=\sum_\beta P_{\alpha\beta}\Phi_\beta^0,\qquad\quad \quad\qquad P_{\alpha\beta}=\sum_i |U_{\alpha i}|^2|U_{\beta i}|^2\label{3ab}.
\end{eqnarray}
Adopting a 3+1 scenario,
$U_{\alpha i}$ denotes a $4\times4$ mixing matrix including a sterile neutrino mixing with
$\nu_\mu$ and $\nu_\tau$. 
The flavor ratios at the detection point are then given by 
($\Phi_{\nu_e}:\Phi_{\nu_\mu}:\Phi_{\nu_\tau}:\Phi_{\nu_s}$), with the neutrino fluxes $\Phi_\beta$ and $\beta\in\{\nu_e,\,\nu_\mu,\,\nu_\tau,\,\nu_s\}$.

For the $3\times3$ sub matrix we adopt an approximate tribimaximal mixing scheme according to \cite{Harrison:2002er}
\begin{eqnarray}
U_{TBM}=\frac{1}{\sqrt{6}}\begin{pmatrix}
                      2&\sqrt{2}&0\\-1&\sqrt{2}&\sqrt{3}\\-1&\sqrt{2}&-\sqrt{3}
                     \end{pmatrix}\qquad\Rightarrow
P_{TBM}=\frac{1}{18}\begin{pmatrix}
                      10&4&4\\4&7&7\\4&7&7
                     \end{pmatrix}\label{z}.
\end{eqnarray}
which, following \cite{Kisslinger:2013sba}, is complemented by the matrices
$O^{e4}, \,O^{\mu4}, \,O^{\tau4}$ modeling the admixture of the sterile neutrino. The rotation matrices $O^{\alpha4}$ are defined as
\begin{align}
 O^{e4}=\begin{pmatrix}
         c_{e}&0&0&s_{e}\\0&1&0&0\\0&0&1&0\\-s_{e}&0&0&c_{e}
        \end{pmatrix}\text{,}\ \ 
O^{\mu4}=\begin{pmatrix}
         1&0&0&0\\0&c_{\mu}&0&s_{\mu}\\0&0&1&0\\0&-s_{\mu}&0&c_{\mu}
        \end{pmatrix}\text{,}\ \  
O^{\tau4}=\begin{pmatrix}
         1&0&0&0\\0&1&0&0\\0&0&c_{\tau}&s_{\tau}\\0&0&-s_{\tau}&c_{\tau}
        \end{pmatrix}\text{,}
\end{align}
with the abbreviations $s_{\alpha}=\sin{\theta_{\alpha}}$ and $c_{\alpha}=\cos{\theta_{\alpha}}$.
This yields the $4\times4$ mixing matrix
$U=U_{TBM}O^{e4}O^{\mu4}O^{\tau4}$,
\begin{eqnarray}
U
=\frac{1}{\sqrt{6}}\left(
\begin{array}{cccc}
 2 c_e & \sqrt{2} c_\mu-2 s_e s_\mu & -s_\tau \left(2 c_\mu s_e+\sqrt{2} s_\mu \right) & c_\tau \left(2 c_\mu s_e+\sqrt{2} s_\mu\right) \\
 -c_e & \sqrt{2} c_\mu+s_e s_\mu & s_\tau (c_\mu s_e-\sqrt{2} s_\mu ) +\sqrt{3} c_\tau & - c_\tau(c_\mu s_e-\sqrt{2} s_\mu)+\sqrt{3} s_\tau \\
 -c_e & \sqrt{2} c_\mu+s_e s_\mu &  s_\tau (c_\mu s_e-\sqrt{2} s_\mu )-\sqrt{3} c_\tau & - c_\tau(c_\mu s_e-\sqrt{2} s_\mu)-\sqrt{3} s_\tau \\
 -\sqrt{6} s_e & -\sqrt{6} c_e s_\mu & -\sqrt{6} c_e c_\mu s_\tau & \sqrt{6} c_e c_\mu c_\tau \\
\end{array} \,\text{,}
\right)\label{3ac}
\end{eqnarray}
For sake of simplicity we assume  either only one mixing angle with the additional state to be non-zero ($s_{\alpha}\equiv\sin\theta$ and $\left.s_{\beta}\right|_{\beta\neq \alpha}=0$) 
or two mixing angles being the same, for instance $s_{\mu}=s_{\tau}\equiv\sin\theta$ and $s_{e}=0$. This allows to calculate the normalized neutrino fluxes resulting from an initial ($\Phi_e^0:\Phi_\mu^0:\Phi_\tau^0:\Phi_s^0$)=($1:2:0:0$) flavor ratio as follows,

\begin{eqnarray}
&&\Phi_e=\frac{1}{3}\Phi_{tot}(P_{ee}+2P_{e\mu})\label{3bb}\text{,}\\
&&\Phi_\mu=\frac{1}{3}\Phi_{tot}(P_{e\mu}+2P_{\mu\mu})\text{,}\\
&&\Phi_\tau=\frac{1}{3}\Phi_{tot}(P_{e\tau}+2P_{\tau\mu})\text{,}\\
&&\Phi_s=\frac{1}{3}\Phi_{tot}(P_{es}+2P_{s\mu})\text{.}
\end{eqnarray}

As a consequence, an 
initial pion source spectrum of 1:2:0 is transformed into the expected flavor ratio of 1:1:1 at the point of detection 
(see e.g. \cite{Learned:1994wg,Ahluwalia:2000fq,Athar:2000yw}).

In order to illustrate the effect of neutrino shortcuts we concentrate on two concrete scenarios for sterile neutrinos:

\begin{itemize}

 \item 
 A scenario with light sterile neutrinos motivated by the LSND, MiniBoone and Gallium anomalies with a $\delta m^2$ of order 1~eV$^2$.
 Here we use
 $\sin^2(2\theta)\simeq0.12$ ($\sin^2\theta\simeq0.03$) and $\delta m^2\simeq1$\,eV$^2$. The parameters used here are compatible with the constraints of $|U_{\mu4}|^2\lesssim 0.03$ and $|U_{\tau4}|^2\lesssim 0.3$ at $90\%$ CL presented in~\cite{Kopp:2013vaa}. Note that to explain the LSND and MiniBoone anomalies, a non-zero $U_{e4}$ mixing is necessary. Here we consider these values for various mixing schemes for illustrative purposes and in particular for the case of $(\nu_\mu,\nu_\tau)-\nu_4$-mixing. For the case of $(\nu_e,\nu_\mu)-\nu_4$ mixing, compare Table \ref{tab1}.
 
\item The neutrino Minimal Standard Model ($\nu$MSM) 
where a keV scale warm dark matter neutrino is introduced. Here we use $\delta m^2\simeq1$\,keV$^2$
and $\sin^2(2\theta)\simeq10^{-7}$ ($\sin^2\theta\simeq10^{-8}$).
\cite{Asaka:2006nq}.
\end{itemize}

For an estimation how large the portion of the neutrino spectrum affected by the resonance is it is helpful to
calculate the Full Width in energy 
at a fraction $f$ of Maximum (FWfM) \cite{Pas:2005rb} as
\begin{equation}
\frac{\delta E({\rm FWfM})}{E_{\rm Res}}=
  \left[1+\,\tan 2\theta\,\sqrt{\frac{1-f}{f}}\;\right]^{1/2}
- \left[1-\,\tan 2\theta\,\sqrt{\frac{1-f}{f}}\;\right]^{1/2}\,,
\end{equation}
which, for small $\theta$ reduces to 
$2\theta\,\sqrt{\frac{1-f}{f}}$.
Thus, the resonance is very narrow for a small mixing angle.
For example, Full Width at Half Max is 
$\delta E({\rm FWHM})= 2\,\theta\, E_{\rm Res}$ for a small angle. 
Assuming a resonance energy of 300~TeV yields 
$\delta E({\rm FWHM})=104$~TeV $(35\%)$, 95 MeV  $(3\cdot 10^{-2}\%)$ for the light sterile and $\nu$MSM scenarios above, respectively.
This demonstrates that the affected portion of the neutrino spectrum depends crucially on the active-sterile neutrino mixing.

\begin{figure}[t]
\centering
\begin{subfigure}
{\includegraphics[width=0.49\textwidth]{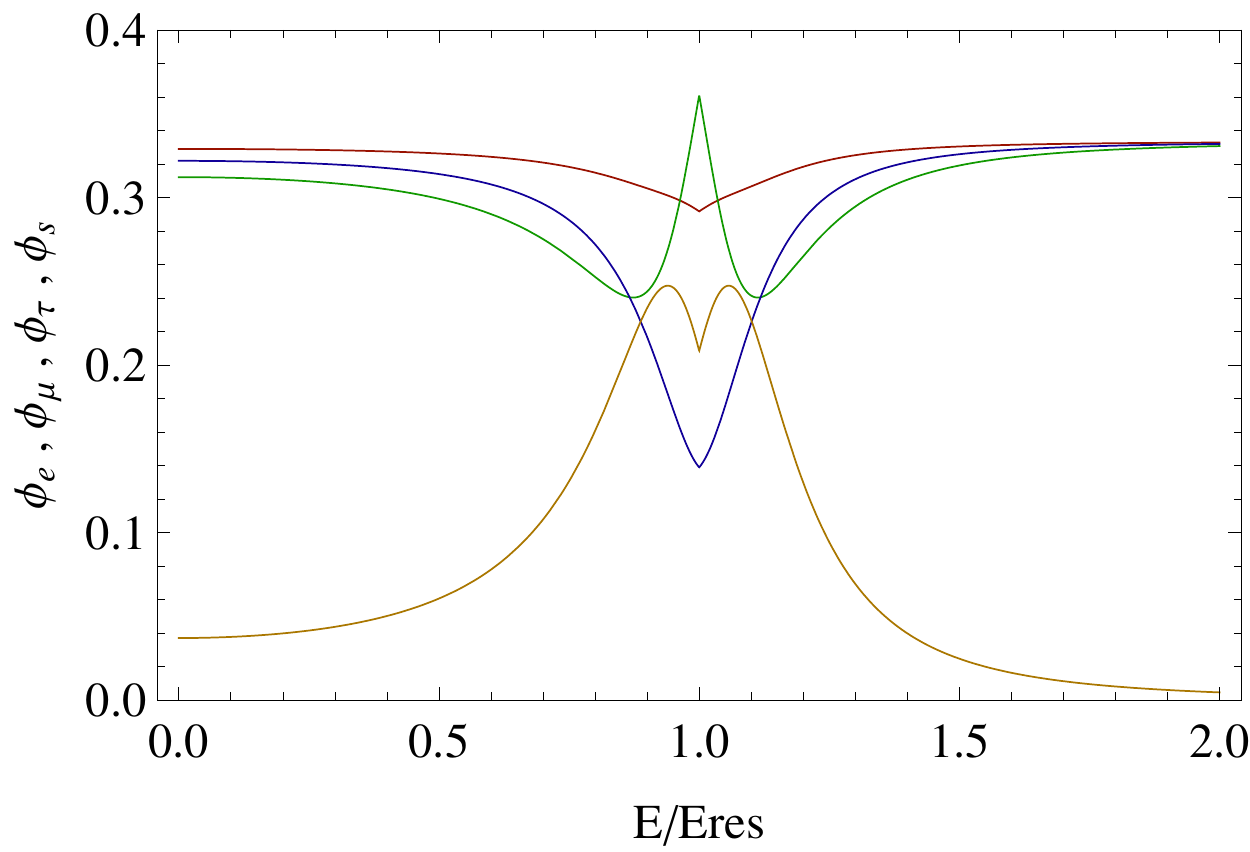}}
\end{subfigure}    
\begin{subfigure}
{\includegraphics[width=0.49\textwidth]{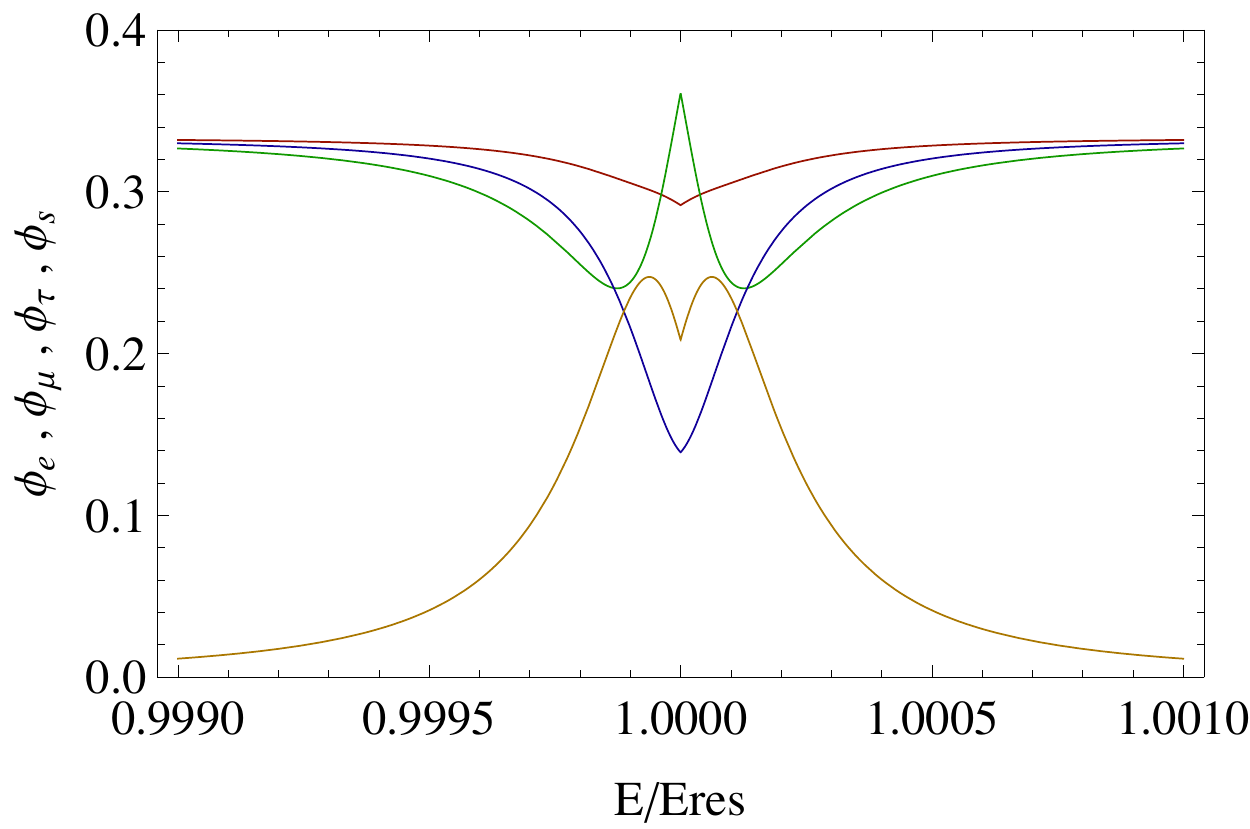}}
\end{subfigure}
\vspace{-0.5 cm}\caption{
Neutrino flavor rates
for a pion source as a function of $E/E_{\rm Res}$:
$\Phi_e$ (red), $\Phi_\mu$ (green), $\Phi_\tau$ (blue) $\Phi_s$ (yellow).
Left panel: light sterile neutrino with $\sin^2(2\theta)=0.12$, $\delta m^2$=1\,eV$^2$.
Right panel: $\nu$MSM  neutrino with $\sin^2(2\theta)=10^{-7}$, $\delta m^2$=1\,keV$^2$.
\label{fig:Bilde3}}
\end{figure}

In figure~\ref{fig:Bilde3} the resonance peaks are shown as a function of $E/E_{\rm Res}$ in the light sterile neutrino (left panel) and
$\nu$MSM scenario (right panel). 

In case of the $\nu$MSM scenario, an energy resolution up to the order of $10^{-2}\%$ is needed to see a substantial modification in the spectrum which is not accessible for current experiments also in view of limited statistics. In the case of the light sterile neutrino scenario and a resonance energy in the 100~TeV region a significant portion of the neutrino spectrum is affected and is expected to be seen with higher statistics in experiments like IceCube.
 
In Table~1 we summarize the change in flavor ratios at resonance energies in the light sterile neutrino for different astrophysical sources and mixing scenarios in addition to the one presented above, the pion source with $(\nu_\mu,\nu_\tau)-\nu_4$ mixing.
The sources are considered to be ideal, based on processes described in \cite{Pakvasa:2007dc}.
Also given are the flavor ratios
$R=\Phi_\mu/(\Phi_e+\Phi_\tau)$, $S=\Phi_e/\Phi_\tau$, $T=\Phi_\mu/\Phi_{\rm tot}$ and $G=\Phi_{\bar{e}}/\Phi_{\rm tot}$, with $\Phi_{\rm tot}=\Phi_{e/\bar{e}}+\Phi_{\mu/\bar{\mu}}+\Phi_{\tau/\bar{\tau}}$.
\newline The breaking of the original $\mu-\tau$ symmetry 
in the fluxes obtained results from
the additional mixing of both flavors with the sterile neutrino. A similar effect due to a non-vanishing $\sin\theta_{13}\neq0$ has been 
found in
\cite{Beacom:2003zg}.  
\newline

\begin{table}[!ht]
\begin{tabular}{l|l||c||c|cc|cc}
source& $\Phi_\beta^0$&mixing&\qquad $\Phi_\beta(\theta^{\alpha 4})$\qquad&\quad$R$\qquad &\quad$S$\qquad&\quad$T$\quad&\qquad$G$\qquad
\\\hline\hline
Pion&1:2:0:0&none ($\theta^{\alpha 4}=0$)&1:1:1:0&1/2&1&1/3&1/9\\
&&$\nu_e-\nu_4$&4:11:11:6&11/19&8/11&11/30&4/45\\
&&$\nu_\mu-\nu_4$&5:5:5:3&1/2&1&1/3&1/9\\
&&$(\nu_e,\nu_\mu)-\nu_4$&32:41:41:30&41/73&32/41&41/114&16/171\\
&&$(\nu_\mu,\nu_\tau)-\nu_4$&21:26:10:15&26/31&21/10&26/57&7/57\\\hline
Damped &0:1:0:0&none ($\theta^{\alpha 4}=0$)&4:7:7:0&7/11&4/7&7/18&2/27\\
Muon &&$\nu_e-\nu_4$& 4:9:9:2&9/13&4/9&9/22&2/33\\
&&$\nu_\mu-\nu_4$&1:2:2:1&2/3&1/2&2/5&1/15\\
&&$(\nu_e,\nu_\mu)-\nu_4$&16:115:115:42&115/131&16/115&115/246&8/369\\
&&$(\nu_\mu,\nu_\tau)-\nu_4$&7:16:4:9&16/11&7/4&16/27&7/81\\\hline
Neutron &1:0:0:0&none ($\theta^{\alpha 4}=0$)&5:2:2:0&2/7&5/2&2/9&5/27\\
Beam &&$\nu_e-\nu_4$&2:1:1:2&1/3&2&1/4&1/2\\
&&$\nu_\mu-\nu_4$&3:1:1:1&1/4&3&1/5&3/5\\
&&$(\nu_e,\nu_\mu)-\nu_4$&10:1:1:6&1/11&10&1/12&5/12\\
&&$(\nu_\mu,\nu_\tau)-\nu_4$&35:14:14:9&2/7&5/2&2/9&5/9
\end{tabular}
\caption{
Difference in flavor rates with $\nu_{\alpha}-\nu_4$ mixing . Here 
$\Phi_\beta^0$ denotes the ratio of initial neutrino rates at the source, ($\Phi_e^0: \Phi_\mu^0: \Phi_\tau^0: \Phi_s^0$), 
$\Phi_\beta(\theta^{\alpha 4}=0)$  denotes the ratio of neutrino rates assuming zero active-sterile neutrino mixing,   ($\Phi_e:\Phi_\mu:\Phi_\tau:\Phi_s$), 
and $\Phi_\beta(\theta^{\alpha 4})$  denotes the ratio of neutrino rates with active-sterile neutrino mixing at resonance energies,
$E=E_{res}$.}
\label{tab1}
\end{table}
\section{MSW-like resonant conversion}
A particularly interesting limit exists in the case where the change in the dispersion relation occurs adiabatically so that an MSW-like
\cite{Wolfenstein:1977ue,Barger:1980tf,Mikheev:1986wj}
resonant conversion can occur
(for details see \cite{PaesSicking}). 
A crucial point in this scenario is that the flavors produced at the source do not propagate as mass eigenstates in vacuum but as eigenstates of the Hamiltonian which have a different admixture of the sterile state due to the change in the dispersion relation. 
Depending on the scenario, a large amount of muon and tau neutrino depletion can be obtained in case of a normal hierarchy if only the heavier mass eigenstates
undergo resonant conversion into sterile neutrinos while the lightest eigenstate does not. In case of inverse hierarchy the same can be achieved if only the heaviest and the lightest mass eigenstates undergo the resonant conversion.
We thus assume a scenario where we
introduce two sterile neutrinos with a small admixture in the mass eigenstates $\nu_2$ and $\nu_3$. 
The mass eigenstate $\nu_1$ which has the maximal contribution of electron neutrinos does not mix with the sterile neutrinos. 
While $\nu_2$ and $\nu_3$ undergo resonant conversion into sterile neutrinos
$\nu_1$ thus will be not affected and the only active component surviving and to be measured. 
Consequently, a total depletion of $\nu_2$ and $\nu_3$
would result in a $U_{e1}^2:U_{\mu 1}^2:U_{\tau 1}^2 \simeq 4:1:1$ flavor ratio above the resonance energy. Moreover, in a scenario where only $\nu_1$ and $\nu_2$ mix with the sterile neutrinos, $\nu_3$ would be the only active component to survive, resulting in a flavor ratio of $U_{e3}^2:U_{\mu 3}^2:U_{\tau 3}^2\simeq 0:1:1$, again above the resonance energy. In both cases large deviations from the expected $1:1:1$ flavor ratio can arise.

Current analyses do not exclude the canonical ratio of 1:1:1 but feature different best fits given as $0:0.2:0.8$ \cite{Aartsen:2015ivb} and $0.92:0.08:0$ \cite{Palomares-Ruiz:2015mka}, depending on the spectral information included in the fit.
The altered flavor ratios arising in the extra-dimensional shortcut scenarios discussed in this paper can at least partly and in specific energy ranges account for
such a depletion of electron or muon and tau neutrino flux. In particular a flavor ratio of $4:1:1$ resembles the best fit in \cite{Palomares-Ruiz:2015mka}. 

For sake of simplicity, in the following we
discuss a 1+1 active-sterile neutrino pattern with a single non-zero mixing angle $\theta$ between an active neutrino $\nu_a$ and a sterile one $\nu_s$. 
As long as the shortcut parameter of an 
asymmetrically warped extra dimension varies adiabatically in this case the neutrino remains in its original eigenstate of
the Hamiltonian. 
If the dispersion relation varies in a way that would generate level crossing in the non-adiabatic regime, the adiabatic limit would lead to resonant conversion. This gives rise to the following probabilities to detect an active or sterile neutrino
in an original beam of pure active neutrinos at its source:
\begin{eqnarray}
  P_{\nu_a\rightarrow\nu_a}(x)=\cos^2\theta\cdot\cos^2\t +\sin^2\theta\cdot\sin^2\t \label{eq_prob} \\
 P_{\nu_a\rightarrow\nu_s}(x)=\sin^2\theta\cdot\cos^2\t +\cos^2\theta\cdot\sin^2\t
\end{eqnarray}
Here $\theta$ denotes the ``vacuum'' mixing angle (i.e. bare angle without ADR effects), whereas $\t$ 
is the effective mixing angle given in equation \eqref{eq_sinq} evaluated at the detection point.
We adopt the metric \eqref{eq_metric} with $\eta^2(u)=\text{e}^{-2k|u|}$. 
This leads to a geodesic oscillating around the brane given by
\begin{equation}
 u(x)=\pm \frac{1}{2k}ln[1+k^2x(l-x)],\quad\quad (\text{with } 0\leq x < l)\text{.} \label{eq_geodesic}
\end{equation}
The shortcut parameter 
\begin{equation}
 \epsilon=1-e^{-k|u|}=1-\frac{1}{\sqrt{1+k^2x(l-x)}}
\label{eq_epsilon}
\end{equation}
now depends on the travel distance, which leads to the effective mixing angle
\begin{equation}
 \cos{2\t}=\frac{\cos{2\theta}-\frac{2E^2}{\delta m^2}\left(1-\frac{1}{\sqrt{1+k^2x(l-x)}}\right)}{\sqrt{\left(\frac{2E^2}{\delta m^2}\left(1-\frac{1}{\sqrt{1+k^2x(l-x)}}
\right)-\cos{2\theta}\right)^2+\sin^2(2\theta)}}\text{.} \label{eq_cost}
\end{equation}
As pointed out in \cite{Hollenberg:2009ws}, the geodesic \eqref{eq_geodesic} repeats itself after $x=l$ to the opposite direction of the brane ($u(x+l)=-u(x)$). This results in a periodicity of $\epsilon$ and therefore of the effective mixing angle $\t$ with the periodic length $l$. It thus is sufficient to use the definition above with $0\leq x < l$. For larger travel distances we exploit the periodic behavior $\cos{2\t}(x)=\cos{2\t}(x+l)$,
see \cite{Hollenberg:2009ws,PaesSicking} for details.
The periodic length $l$ is a function of the initial conditions for the velocities 
and of the warp factor $k$
\cite{Hollenberg:2009ws}:
\begin{equation}
 l=2\frac{|\dot{u_0}|}{k\dot{x_0}},
\end{equation}
where $\dot{x_0}$ and $\dot{u_0}$ denote the initial speed of the sterile neutrino along and perpendicular to the brane, respectively. 

The limit of resonant conversion applies as long as the change in the dispersion relation is adiabatic, corresponding to the following condition:
\begin{align}
 \gamma_{res}=\frac{4E^3}{(\delta m^2)^2\sin^2(2\theta)}\left. \frac{d\epsilon}{dx}\right|_{res}\ll 1
\end{align}
where the shortcut parameter $\epsilon$ is given by \eqref{eq_epsilon}.
This is the condition for adiabatic transition at resonance. 
If the condition is fulfilled at resonance, it is fulfilled everywhere else. 
As the derivative of the shortcut parameter $\epsilon$ with respect to the baseline $x$ at resonance becomes rather complicated, 
for sake of simplicity
we adopt the maximum value of $\frac{d\epsilon}{dx}$ at $x=0$ as an upper bound:
\begin{align}
  \left. \frac{d\epsilon}{dx}\right|_{res}<\left. \ \frac{d\epsilon}{dx}\right|_{max}=\frac{1}{2} k^2 l.
\end{align}
Consequently, if the condition
\begin{align}
 \gamma_{max}=\frac{2E^3}{(\delta m^2)^2 \sin^2(2\theta)}k^2l\ll 1
\end{align}
is fulfilled, the adiabatic condition is fulfilled as well.

\begin{figure}[t]
\includegraphics[width=.45\textwidth]{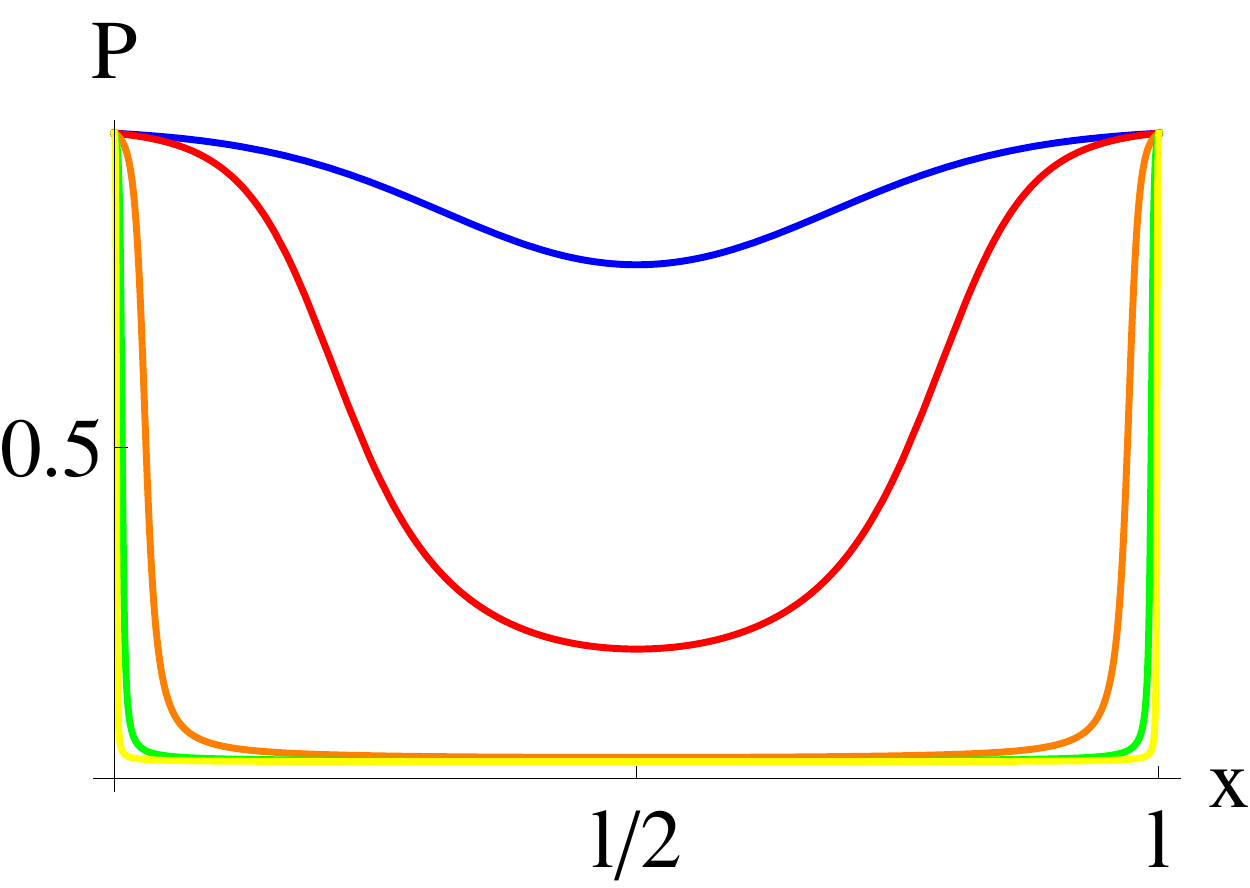}
\includegraphics[width=.45\textwidth]{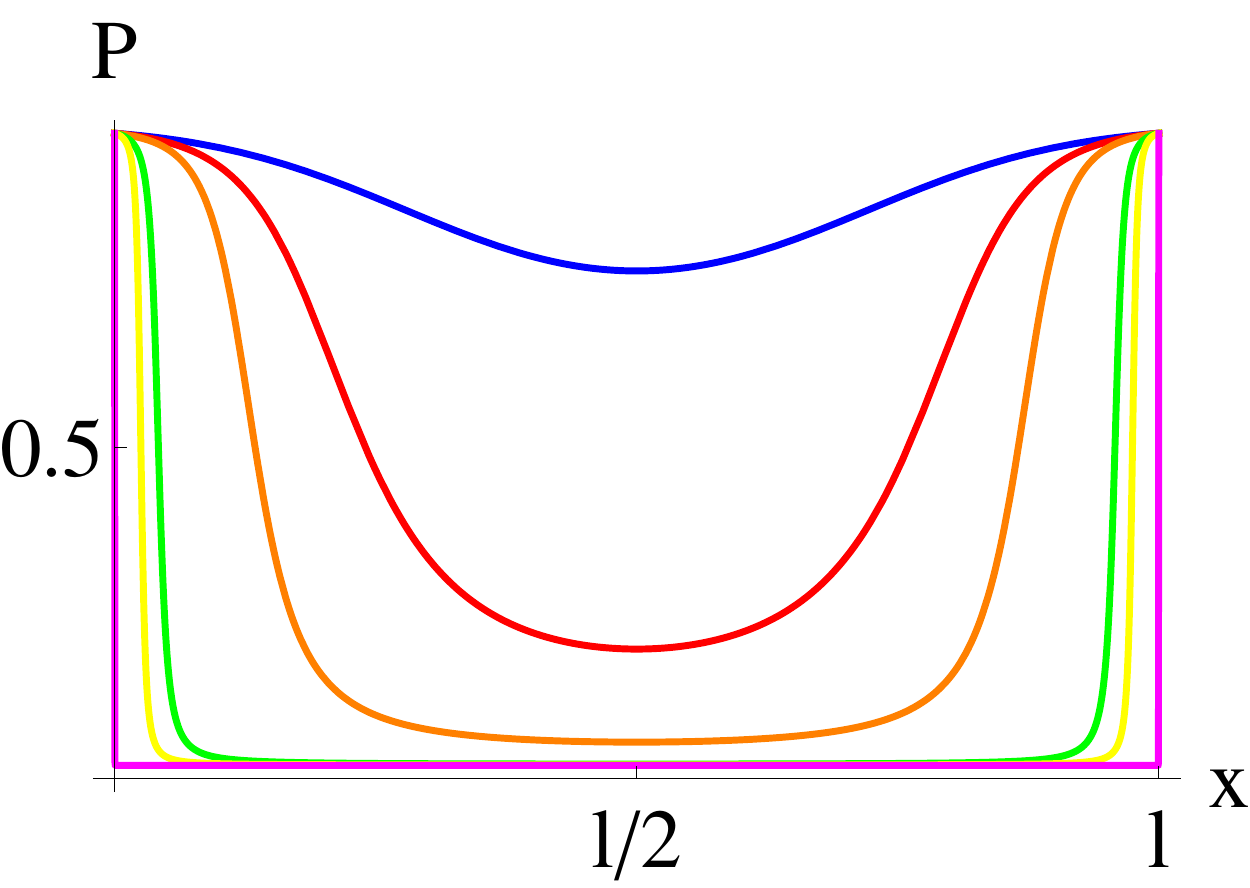}
\caption{Survival probability \eqref{eq_prob} as a function of the travel distance $x$ for various geometric factors $k$
(left panel, blue: $k=3/l$, red: $k=4/l$, orange: $k=10/l$, green: $k=20/l$, yellow: $k=40/l$;
adopting $\frac{\delta m^2}{E^2}=1$ to illustrate the dependence on $k$) and energies $E$ (right panel,
blue: $E^2=0.8\cdot \delta m^2$, red: $E^2=1\cdot \delta m^2$, orange: $E^2=1.2\cdot \delta m^2$, green: $E^2=1.4\cdot \delta m^2$, yellow: $E^2=2.3\cdot \delta m^2$, purple: $E^2=3.2\cdot\delta m^2$; adopting a fixed value of $k=4/l$).
The vacuum mixing angle is fixed to $\cos{2\theta}=0.99$ in both panels. Due to the periodic behaviour of \eqref{eq_cost}, the shown survival probability will repeat itself for $x>l$.
(Note that the parameters in this plot have been chosen to illustrate the effect. Realistic parameter ranges will be discussed below.)}
\label{fig_survprob}
\end{figure}

As can be seen in \eqref{eq_cost} the survival probability depends on the travel distance of the neutrinos. The survival probability is shown in Figure \ref{fig_survprob} for different values of $k$ and different energies. For larger values of $k$ the probability becomes steeper in the region of $x=0,l$, but is nearly constant at its minimal value $P_{min}=\frac12(1-\cos(2\theta))$ for all travel distances, which are longer than the resonance length.
In the following we distinguish two cases: First, the periodic length is larger than the travel distance. For this case it is crucial that the resonance crossing occurs during the propagation distance, which is satisfied if the resonance length 
\begin{equation}
 x_{res}=\frac l2 -\sqrt{\frac{l^2}{4}+\frac{1}{k^2}\left(1-\frac{1}{1-\frac{\delta m^2 \cos{2\theta}}{2E^2}}\right)}
\approx\frac{1}{k^2l}\frac{\delta m^2 \cos{2\theta}}{2E^2} 
\label{eq_resonanz}
\end{equation}
is much smaller than the diameter of the Milky Way. This leads to
\begin{align}
x_{res}<10^6 \text{ly}=5\cdot 10^{25}\frac1{\text{eV}}\\
\Rightarrow k^2l > \frac{1}{5} 10^{-25} \text{eV} \frac{\delta m^2 \cos{2\theta}}{2E^2}\text{.}
\end{align}
If this condition is satisfied the survival probability becomes $P_{min}$, which is nearly negligible due to the small vacuum mixing angle $\theta$. The approximation in equation \eqref{eq_resonanz} is valid if the second term in the square root is small compared to the first one, which leads to the condition
\begin{align}
 \frac{\delta m^2 \cos{2\theta}}{2E^2} + \frac{l^2k^2}{4}\gg\frac{\delta m^2 \cos{2\theta}}{2E^2}\cdot\frac{l^2k^2}{4}\text{.}
\end{align}
This condition holds in our context since we only consider high energies and therefore $\frac{\delta m^2 \cos{2\theta}}{2E^2}\ll 1$.

\begin{figure}[t]
\includegraphics[width=.45\textwidth]{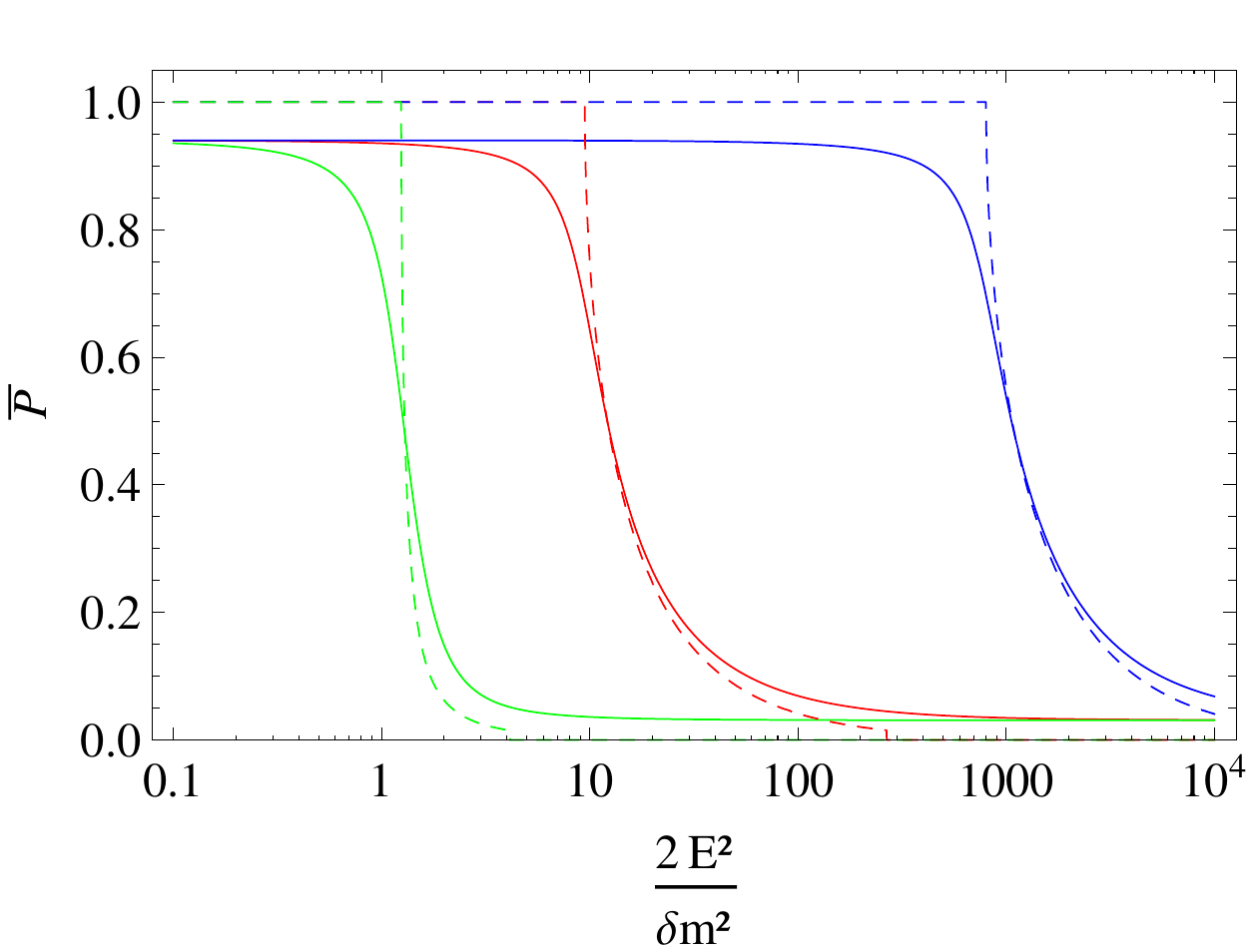}
\caption{
Averaged probability of measuring active neutrino in dependence of the
energy-scale $\frac{2E^2}{\delta m^2}$. The different colors correspond
to different values of the product $kl$: Blue: $kl=0.1$, Red: $kl=1$,
Green: $kl=10$. The solid lines refer to the light sterile neutrino model with
$\sin^2(2\theta)=0.12$ while the dashed lines refer to the
$\nu$MSM-Model with $\sin^2(2\theta)=10^{-7}$.
(Note that the parameters in this plot have been chosen to illustrate the effect.
Realistic parameter ranges will be discussed below.)}
\label{fig_AveKL}
\end{figure}

Second, the periodic length is smaller than the typical travel distance. Due to the assumption that the events measured by IceCube do not have one common origin, the neutrinos arriving at the detector do so after having propagated over different travel distances. Note that
when the travel distances are larger than the periodic length, periodicity implies that averaging over cosmic sources with $L\gg l$ is equivalent
to averaging over $l$.
Therefore we average over the periodic length to get an averaged survival probability 
\begin{equation}
\bar{P}=\frac{1}{l}\int_0^l P_{\nu_a\rightarrow\nu_a}(x) dx=\int_0^l\left[\cos^2\theta\cdot\cos^2\t(x) +\sin^2\theta\cdot\sin^2\t(x)\right]dx
\end{equation}
which with the help of  \eqref{eq_cost} reduces to
\begin{eqnarray}
 \bar{P}=\frac12(I\cos{2\theta}+1)
\end{eqnarray}
where
\begin{equation}
I=\frac{1}{l}\int_0^l \frac{\cos{2\theta}-\frac{2E^2}{\delta m^2}\left(1-\frac{1}{\sqrt{1+k^2x(l-x)}}\right)}{\sqrt{\left(\frac{2E^2}{\delta m^2}\left(1-\frac{1}{\sqrt{1+k^2x(l-x)}}
\right)-\cos{2\theta}\right)^2+\sin^2(2\theta)}} dx\text{.}
\end{equation}
This expression yields an energy dependent transition probability. As can seen in figure \ref{fig_AveKL}, the averaged survival probability
breaks down at the resonance energy $E_{\rm Res}$ and is extremely low for higher energies. This corresponds to the fact, that 
the effect of shortcuts is only relevant at higher energies. For lower energies there is no significant deviation to standard neutrino mixing.
Assuming an adiabatic change in the dispersion relation
the high and low energy limits for the transition and survival probabilities can be calculated as follows,
\begin{align}
 \bar{P}(E=0)=\frac{1}{2}(\cos^2{2\theta}+1)\\
 \bar{P}(E\rightarrow\infty)=\frac{1}{2}(1-\cos(2\theta))\text{.}
\end{align}
 The survival probability $\bar{P}$ depends only on the product of the geometric parameters $k$ and $l$. 
As can be seen the fall-off of the survival probability is shifted
to higher energies for smaller factors $kl$. This corresponds to a smaller ratio of the initial velocities $\frac{|\dot{u_0}|}{\dot{x_0}}$. 
For smaller $kl$
the sterile neutrino thus plunges less deeply into the bulk
and thus is less affected by the shortcut, therefore it needs a 
have higher energy to experience resonant conversion. 

Fixing the product $kl$ and varying the vacuum mixing angle does not affect the
position of the tilt in the conversion probability  
but makes the function steeper 
for smaller vacuum mixing angles. 

For a given set of mixing parameters $\delta m^2$, $\sin^2(2\theta)$ and a given energy $E$ we obtain the following constraints 
for the geometric parameters $k$ and $l$:

If the product $kl$ is too small, no level crossing will occur in the non adiabatic region, and consequently no resonant conversion will emerge.
On the other hand, if
the product $k^2l$ is too large, the adiabatic condition will not be fulfilled and the level crossing will not result in resonant conversion. Additionally for large periodic length $l$ there is a lower bound for the product $k^2l$ which ensures that the resonance length is much smaller than the expected travel distance.
Adopting again the parameter sets for light sterile and $\nu$MSM neutrinos discussed above, and requiring a transition probability 
$>75~\%$ we arrive at the conditions under which the major part of $\nu_2$ and $\nu_3$ and therefore a large part of muon and tau neutrinos are converted into sterile neutrinos.
Numerical values for various parameter sets and neutrino energies are given in table~\ref{tabbound}.

\begin{table}[t]
\begin{tabular}{c|r|r|r|r|r|r}
\multicolumn{1}{c|}{Model}& \multicolumn{1}{c|}{$\delta m^2$}& \multicolumn{1}{c|}{$\sin^2(2\theta)$}& 
\multicolumn{1}{c|}{$E$}& \multicolumn{1}{c|}{$kl$}& \multicolumn{1}{c|}{$k^2l$/eV}& \multicolumn{1}{c}{$\frac{\delta m^2\cos{2\theta}}{E^2}$}\\\hline
light sterile neutrino & (1eV)$^2$& 0.12 & 30 TeV & $\gtrsim 10^{-13}$& $\ll2.2\cdot10^{-42}$& $10^{-27}$\\
light sterile neutrino  & (1eV)$^2$& 0.12 & 1.2 PeV & $\gtrsim 10^{-15}$& $\ll3.5\cdot10^{-47}$& $6.5\cdot10^{-31}$\\
$\nu$MSM & (1 keV)$^2$& $10^{-7}$ & 30 TeV & $\gtrsim 10^{-11}$& $\ll5.5\cdot10^{-29}$& $1.1\cdot10^{-21}$\\
$\nu$MSM & (1 keV)$^2$& $10^{-7}$ & 1.2 PeV & $\gtrsim 10^{-13}$& $\ll3.5\cdot10^{-32}$&  $6.9\cdot10^{-25}$\\
\end{tabular}

\caption{Regions in the parameter space of  $k$ and $l$ where resonant conversion occurs in various parameter sets and neutrino energies $E$.
\label{tabbound}
} 
\end{table}

\begin{figure}[t]
\includegraphics[width=.85\textwidth]{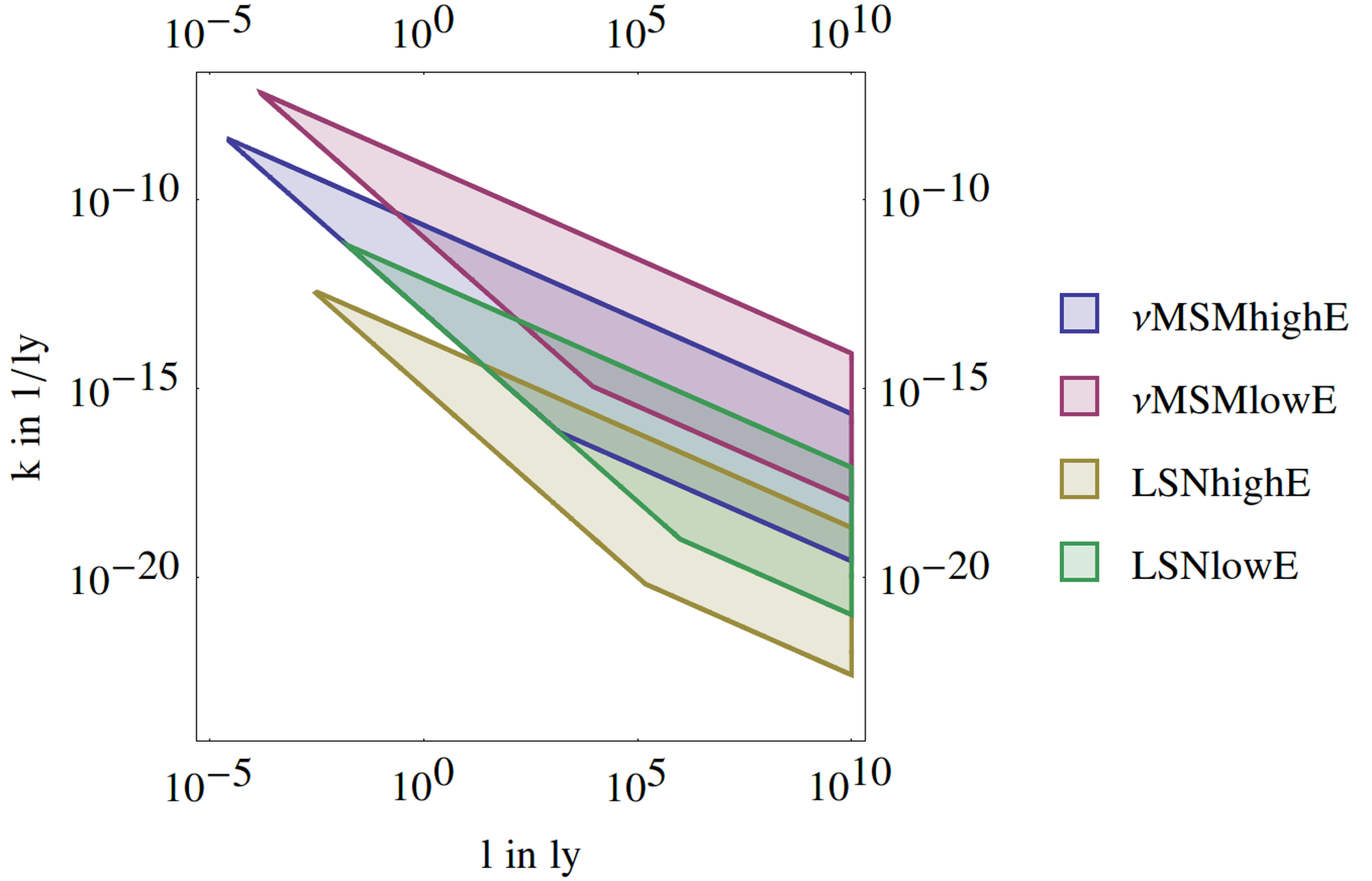}
\caption{Regions in the parameter space of  $k$ and $l$ where resonant conversion occurs. Bounds on the resonance length from atmospheric/solar neutrino oscillations have been taken into account. The different cases depicted correspond to a $\nu$MSM neutrino and a light sterile neutrino (LSN) with energies of 30~TeV and 1.2~PeV.}
\label{fig_bound}
\end{figure}

\begin{figure}[t]
\includegraphics[width=.85\textwidth]{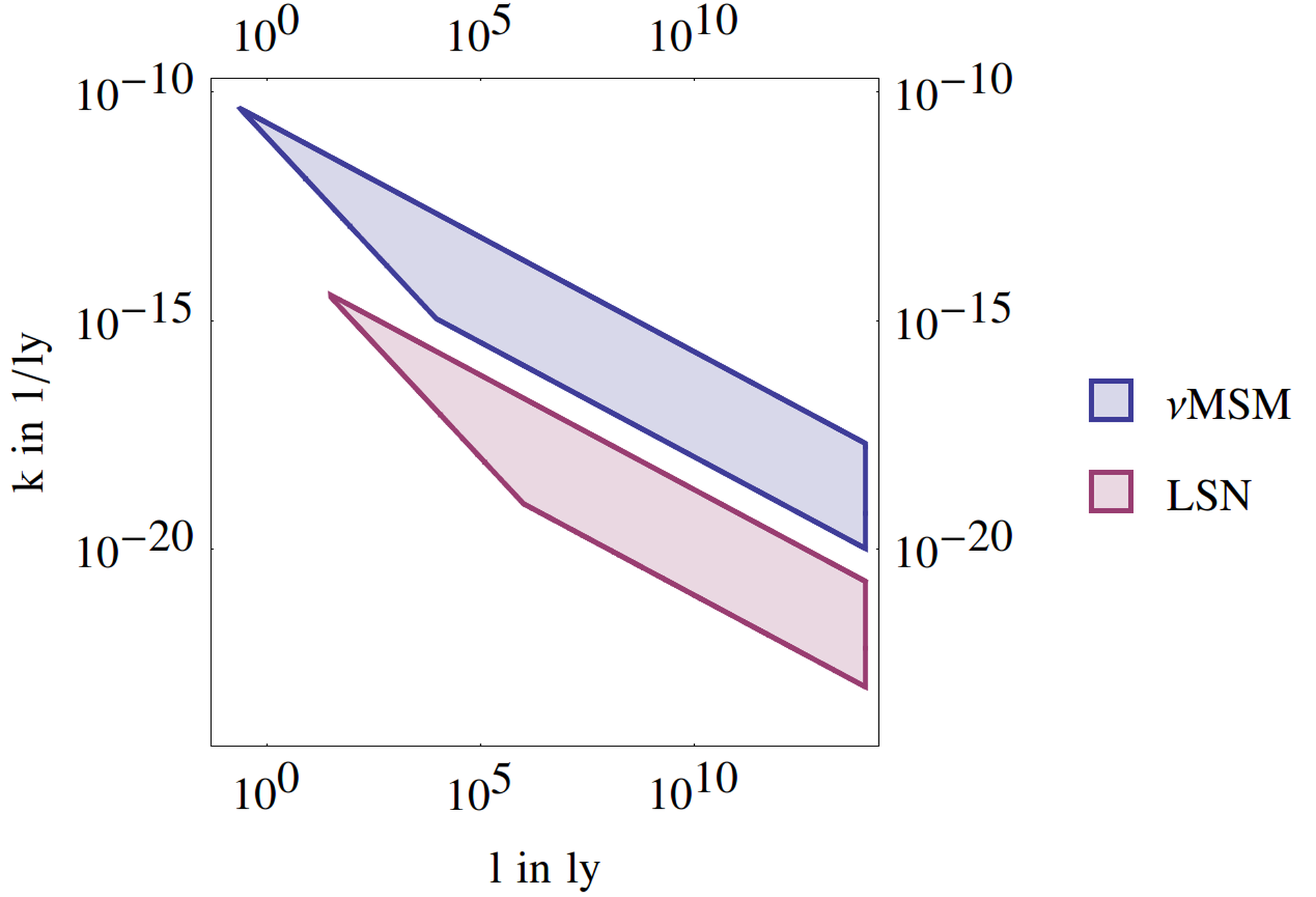}
\caption{As Fig. \protect{\ref{fig_bound}}, but displaying the regions in parameter space where
resonant conversion occurs in the entire interval from
30~TeV to 1.2~PeV.}
\label{fig_allbounds}
\end{figure}

Figure ~\ref{fig_bound} displays the bounds obtained on $k$ and $l$.
As can be seen the periodic length $l$ has to be very large to fulfill both conditions.
Additionally figure \ref{fig_allbounds} shows the allowed regions for the geometric parameters when the whole energy region from 30~TeV up to 1.2~PeV should be affected.
Also taken into account is the constraint 
resulting from the fact that
atmospheric and solar neutrino data are in conflict with significant oscillation or conversion into sterile
neutrinos. In order to have atmospheric and solar neutrinos unaffected by the change in the dispersion relation  
we assume that the resonance length is much larger than the distance from the Earth to the sun
\begin{equation}
x_{res}>8\cdot 10^{17}\frac1{\text{eV}}=1.6\cdot 10^{-5}\text{ly}\label{eq_milky}
\end{equation}
which leads to
\begin{equation}
 k^2l<\frac{1}{8}10^{-17} \text{eV} \frac{\delta m^2 \cos{2\theta}}{2E^2}  \text{.}
\end{equation}
As can be seen in figures \ref{fig_bound} and \ref{fig_allbounds} large parts of the parameter space are consistent with all constraints and 
allow for a significant reduction of the rates of astrophysical muon and tau neutrinos. 

\section{Conclusion}
In summary, we have discussed the effect of altered dispersion relations for sterile neutrinos propagating in extra dimensions  on 
the flavor ratios of astrophysical neutrinos. We have confined our discussion to two important limits: the case where the extra-dimensional
shortcut can be described in terms of a constant effective potential and the case where the potential changes adiabatically giving rise to
an MSW-like resonant flavor conversion. In the first case one gets significant distortions of the expected flavor ratio around the resonance
region whose width depends strongly on the active-sterile neutrino mixing. In the second case strong deviations, including a significant suppression
of muon and tau neutrino fluxes are possible over a large range of energies and baselines. This effect arises only for energies above the resonance energy and may be a possible explanation for the unexpected energy dependence reported in \cite{Palomares-Ruiz:2015mka}.  The flavor ratios of high energy astrophysical neutrinos thus qualify as
a prime probe for non-standard dispersion relations originating from exotic physics such as shortcuts in extra dimensions.

\acknowledgments
We thank  Danny Marfatia and our first referee for useful discussions.
HP was supported by the 'Helmholtz Alliance for Astroparticle Physics HAP' funded by the Initiative and Networking Fund of the Helmholtz 
Association. SP was supported by the US DOE
grant DE-Fc02-04ER41201 and the Alexander von Humboldt Foundation and 
acknowledges TU Dortmund for kind hospitality.
HP and SP thank the KITP Santa Barbara for kind hospitality. This research was supported in part by the National Science 
Foundation under Grant No. NSF PHY11-25915.

\end{document}